\documentclass[a4paper,fleqn]{cas-sc}

\usepackage{hyperref}

\usepackage[T1]{fontenc}
\usepackage{charter}
\usepackage[numbers]{natbib}
\usepackage{todonotes}
\usepackage{listings}
\usepackage{url}
\hypersetup{
    colorlinks=true,                
    breaklinks=true,                
    urlcolor= blue,                
    linkcolor= blue,                
    bookmarksopen=false,
    filecolor=black,
    citecolor=blue,
    linkbordercolor=blue
}

\usepackage{listings}
\usepackage{cleveref}

\def\tsc#1{\csdef{#1}{\textsc{\lowercase{#1}}\xspace}}
\tsc{WGM}
\tsc{QE}
\tsc{EP}
\tsc{PMS}
\tsc{BEC}
\tsc{DE}

\begin{document}

\shorttitle{EATXT: A textual concrete syntax for EAST-ADL}

\shortauthors{Zhang et~al.}

\title [mode = title]{EATXT: A textual concrete syntax for EAST-ADL \\ {\Large An exemplar for blended modeling research}}  



%
\author[1]{Weixing Zhang}[type=editor,
                        orcid=0000-0003-2890-6034]



\ead{weixing.zhang@gu.se}



\affiliation[1]{organization={Chalmers $|$ University of Gothenburg},
    city={Gothenburg},
    country={Sweden}}

\author[2]{Jörg Holtmann}
\author[1,3]{Daniel Strüber}
\author[4]{Jan-Philipp Steghöfer}





\affiliation[2]{organization={Independent Researcher},
    city={Paderborn},
    country={Germany}}

\affiliation[3]{organization={Radboud University},
    city={Nijmegen},
    country={Netherlands}}
    
\affiliation[4]{organization={Xitaso GmbH IT \& Software Solutions},
    city={Augsburg},
    country={Germany}}

\cortext[cor1]{Corresponding author}



\begin{abstract}
Blended modeling is an approach to enable users to interact with a model via multiple notations.     
In this context, there is a growing need for open-source industry-grade exemplars of languages with available language engineering artifacts, in particular, editors and notations for supporting the creation of models based on a single metamodel in different representations (e.g., textual, graphical, and tabular ones). These exemplars can be used to support the development of advanced solutions to address the practical challenges posed by blended modeling requirements.

As one such exemplar, this  paper introduces EATXT, a textual concrete syntax for automotive architecture modeling with EAST-ADL, developed in cooperation with an industry partner in the automotive domain. 
The EATXT editor is based on Xtext and provides basic and advanced features, such as an improved content-assist and serialization specifically addressing blended modeling requirements. We present the editor features and architecture, the implementation approach, and previous use of EATXT in research.
The EATXT editor is publicly available, rendering it a valuable resource for language developers. 
\end{abstract}


\maketitle

\section{Introduction}
\label{sec:intro}
\textit{Blended modeling} is an emerging modeling technique that involves seamless interaction among multiple notations (i.e., concrete syntaxes) and a single model (i.e., abstract syntax)~\cite{ciccozzi2019blended, david2023blended}. It enhances the modeling flexibility of tools and the productivity of engineers, who can choose an appropriate syntax according to their preferences and the task at hand. 
Yet, blended modeling also leads to new challenges, as it is hard to keep models and different language definition artifacts synchronized with each other, even more so when the language evolves. 
To inform the design and evaluation of such approaches, there is a need for open-source industry-grade exemplars of languages with metamodels and associated language engineering artifacts.

The contribution of this paper is such an exemplar for the scope of  EAST-ADL modeling. EAST-ADL is a widely used domain-specific language used for automotive embedded systems~\cite{eadl}.
First steps towards blended modeling were enabled by a previous EAST-ADL implementation called EATOP~\cite{eatop} that supports tabular concrete syntaxes, a tree-based editor, and, in some customized and non-public versions, a graphical  syntax~\cite{yuan2014safety}. However, it lacks support for a textual concrete syntax. This limitation hinders the advantages of textual modeling when using EATOP for EAST-ADL modeling, e.g., text batch replacement, convenient keyboard-based editing, simple searching, and support for text-based copy-and-paste. Lacking a textual syntax affects the maximization of engineer throughput and modeling flexibility.

In this paper, we present EATXT, a concrete syntax for textual EAST-ADL  modeling.
We developed EATXT to address blended modeling requirements, in cooperation with an industry partner from the automotive domain. 
EATXT includes the definition of the syntax as well as a textual editor that supports it.
This concrete syntax is based on the metamodel of EAST-ADL, which represents all domain concepts of EAST-ADL and the relationships between them. The textual editor that supports EATXT's concrete syntax provides common features of modern editors such as content-assist, auto-completion, grammar highlighting, and automatic formatting, etc. 
EATXT serves as an exemplar of an industry-level open-source textual syntax and editor that can be used to support the development of language engineering approaches in the context of blended modeling.
In our online publication of the associated software artifact \cite{eatxtarchive},  we have followed artifact-sharing guidelines~\cite{damasceno2021quality} for ensuring artifact quality and reuse potential.



The remaining sections of this paper are organized as follows. Section~\ref{sec:background} describes the background of this research.  Section~\ref{sec:example} presents some usage scenarios. Section~\ref{sec:impl} outlines the architecture, implementation approach, and features of EATXT. Section~\ref{sec:previous_use} presents the use of EATXT in previous research. Section~\ref{sec:conclusion} summarizes the conclusions and outlines future work.

\section{Background}
\label{sec:background}

\subsection{EAST-ADL and EATOP}\label{sec:background_EASTADL}
EAST-ADL is a domain-specific language for describing automotive embedded system architectures~\cite{cuenot2007managing}. It is based on a large metamodel with 291 metaclasses~\cite{zhang2024supporting} and a nested element hierarchy describing different aspects of electronic vehicle systems. It allows comprehensive system modeling of automotive embedded systems with early analysis capabilities. It is managed by the EAST-ADL Association~\cite{eadl} and developed by different projects and stakeholders. The association has released several versions EAST-ADL versions.

Eclipse EATOP is an infrastructure platform implementation of common base functionality for design tools that provide features to edit EAST-ADL models.
EATOP is an open-source project under the Eclipse Modeling project~\cite{eatop}. Hence, it provides Eclipse-based user interface support, encompassing features such as wizards for creating EAST-ADL projects,
preferences for EAST-ADL versions, property pages for EAST-ADL projects, EATOP perspectives, a resource explorer view, a tree-based editor, etc. EATOP includes implementations of the metamodel for two versions of EAST-ADL. 
EATOP can be employed as an external tool or installed as standalone software within the Eclipse IDE environment. Additionally, a set of design tools 
is available for manipulating EAST-ADL model instances. 

EATOP applies a custom persistence format
called EAXML. 
EAXML is a customized variant of the conventional EMF persistence format XMI and, beyond custom XML tags, preserves the order of the persisted elements according to their representation in the tree-based editor.

\subsection{Developing metamodel-based textual languages with Xtext}
Xtext~\cite{Xtext} is a framework for developing programming languages and domain-specific languages (DSLs)~\cite{behrens2008xtext}. While a variety of tools for the same purpose exist~\cite{erdweg2015evaluating}, Xtext is particularly widely used in the model-driven enginering community due to its native integration with the Eclipse ecosystem. There are two basic concepts and artifacts in Xtext, namely grammar and metamodel. Xtext supports generating Xtext artifacts from grammars, which constitutes the infrastructure for a textual editor. Grammars can be defined manually directly or generated from metamodels. In the latter case, users first represent domain concepts and their relationships by creating metamodels. A grammar is generated from this metamodel and subsequently a text editor is generated from this grammar. The generation process is controlled by the Modeling Workflow Engine (MWE2)~\cite{mwe}. Running MWE2 generates a 
infrastructure, including a parser, linker, type checker, compiler, and editing support for Eclipse, any editor that supports the language server protocol, and a web browser.

\section{Features of the EATXT Editor}
\label{sec:example}
\begin{figure}[t]
    \centering
    \includegraphics[width=\textwidth]{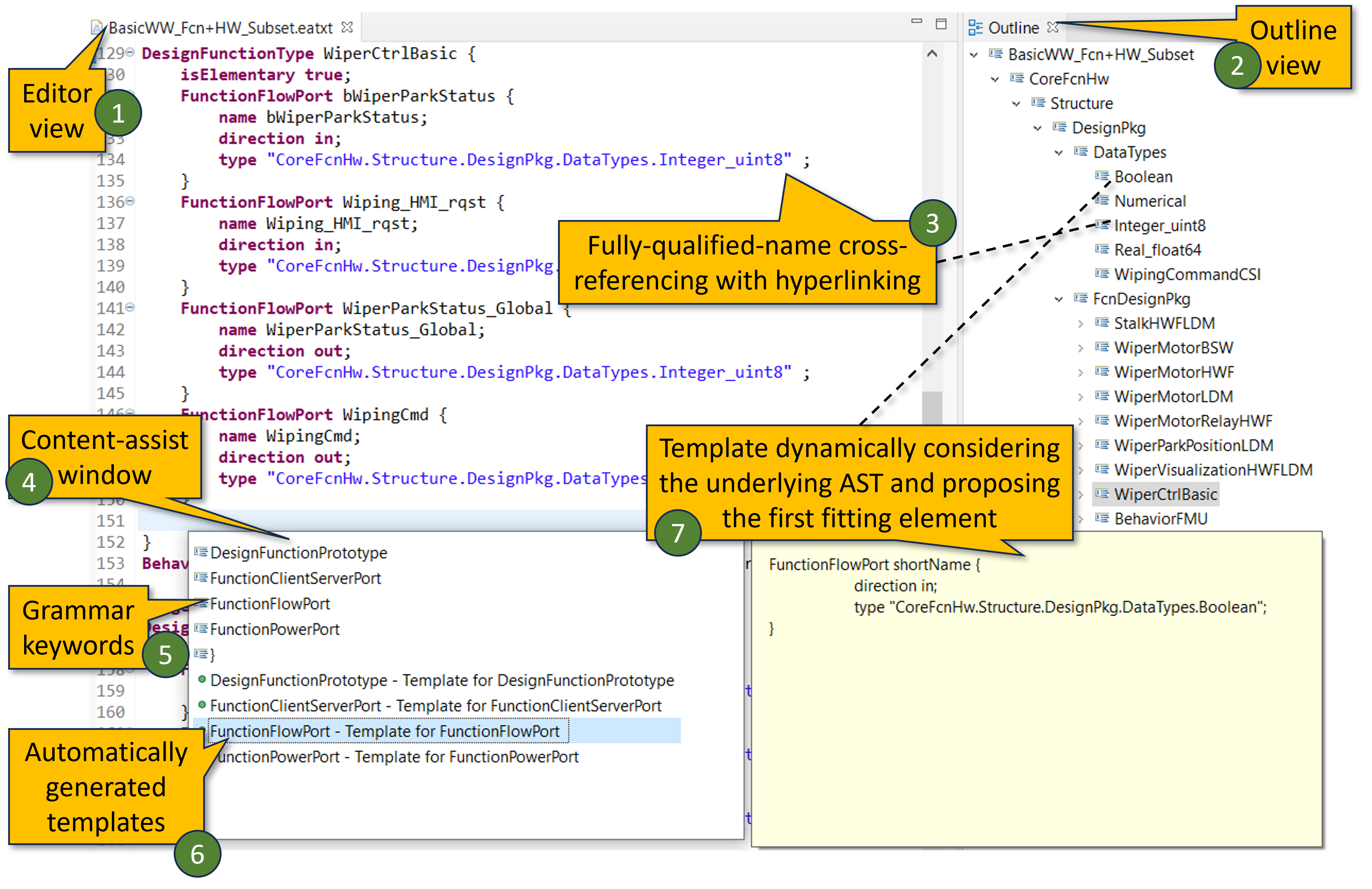}
    \caption{An overview of the EATXT editor.}
    \label{fig:EATXT_example}
\end{figure}

In this section, we present the EATXT editor in terms of its user-facing features. 
These features rely on, but go significantly go beyond the standard functionalities of editors automatically generated by Xtext.
For the purpose of illustration, \Cref{fig:EATXT_example,fig:Serialization_example} show screenshots of the EATXT editor with annotations that we explain throughout this section.
The figures depict excerpts of an example model that specifies an  automotive wiper control unit.
As EAST-ADL is an architecture description language, it entails typical architectural language concepts such as component-like elements with ports. 



\subsection{Editor and outline view}\label{sec:example_views}
Xtext and the underlying Eclipse framework provide different views on a loaded resource by default. 
In the context of the EATXT editor, the user loads a persisted textual model in the form of a file resource with the suffix ``.eatxt'', so that the corresponding resource gets displayed in the different views. 

\Cref{fig:EATXT_example} illustrates this with the loaded file resource \textsf{BasicWW\_Fcn\_Subset.eatxt}.
The editor view (annotation \textsf{1}) shows and enables editing the concrete textual syntax of EAST-ADL models.
Container model elements are represented as text blocks delimited by curly braces and prefixed by the corresponding EAST-ADL metaclass name (e.g., the \textsf{DesignFunctionType} \texttt{WiperCtrlBasic} starting at line 129 in \Cref{fig:EATXT_example}).
Members of the model elements can be further container elements (e.g., the property \textsf{FunctionFlowPort} \texttt{bWiperParkStatus} starting at line 131), attributes, or cross-references to other elements.
Attributes are represented by the corresponding attribute name and value (e.g., the attribute \textsf{isElementary} with the value \texttt{true} in line 130).
Cross-references are represented by their name and the path to the corresponding target (e.g., the cross-reference \textsf{type} of the \textsf{FunctionFlowPort} \texttt{bWiperParkStatus} in line 134).

Complementary, the outline view (annotation \textsf{2}) provides an overview of the overall EATXT model as well as quick navigation to its elements.
Furthermore, it highlights in which context the user is editing in the editor view (see the highlighting of the EAST-ADL model element \texttt{WiperCtrlBasic}).

As an addition to these standard functionalities, we implemented cross-references to always be rendered with their fully qualified names, even in the same container \cite{xtextautomation2023}.
This feature eases the serialization to the native EAST-ADL persistence format EAXML (cf.~\Cref{sec:background_EASTADL}), which we exemplify in \Cref{sec:example_EAXML}.
This aspect is illustrated in \Cref{fig:EATXT_example} by annotation \textsf{3}, where the attribute \textsf{type} of the \textsf{FunctionFlowPort} \texttt{bWiperParkStatus} points to its fully qualified name in the overall EAST-ADL model.
Xtext and Eclipse provide the functionality to navigate directly to the corresponding model element 
via an editor-internal hyperlink.
\subsection{Content-assist}\label{sec:example_contentassist}
The annotation \textsf{4} in \Cref{fig:EATXT_example} points to the content-assist window of EATXT.
Xtext and Eclipse provide the default functionality of providing content-assist proposals that are sensitive to the current context from which the content-assist was called.
In \Cref{fig:EATXT_example}, the content-assist was triggered from line 151, which resides in the model element container \texttt{WiperCtrlBasic} typed by the EAST-ADL metaclass \textsf{DesignFunctionType}.
Accordingly, the content-assist provides proposals for all members that can be added to a \textsf{DesignFunctionType}.

The upper elements of the content-assist window (annotation \textsf{5}) provide plain grammar keywords for such member elements.
For example, the user can add further \textsf{FunctionFlowPort}s to \texttt{WiperCtrlBasic}. 
If the user chooses such a content-assist proposal, only this grammar keyword is auto-completed (i.e., the String ``\textsf{FunctionFlowPort}'').

In contrast, the lower elements of the content-assist window (annotation \textsf{6}) provide complete text snippets that propose a full model element with certain blanks that have to be filled by the user.
Xtext and Eclipse call such complete text snippets \emph{templates}.
Annotation \textsf{7} points to the template proposal for a new \textsf{FunctionFlowPort}.
If the user chooses such a content-assist proposal, the full text snippet is auto-completed.
We extended the template mechanism in such a way that EATXT proposes as few as possible blank lines in the template \cite{xtextautomation2023}.
To achieve this, we on the one hand exploit the EAST-ADL metamodel structure to let EATXT auto-complete all mandatory metamodel elements (e.g., in annotation \textsf{7}, for \textsf{FunctionFlowPort} the members \textsf{direction} and \textsf{type} are mandatory).
On the other hand, we exploit the EAST-ADL metamodel structure to let EATXT consider the current Abstract Syntax Tree (AST) of the EAST-ADL model and dynamically propose the first fitting element of model references (e.g., EATXT proposes the model element \texttt{CoreFcnHw\dots Datatypes.Boolean} for the reference \textsf{type}). 

\subsection{Serialization of EAXML}\label{sec:example_EAXML}

\begin{figure}[t]
    \centering
    \includegraphics[width=\textwidth]{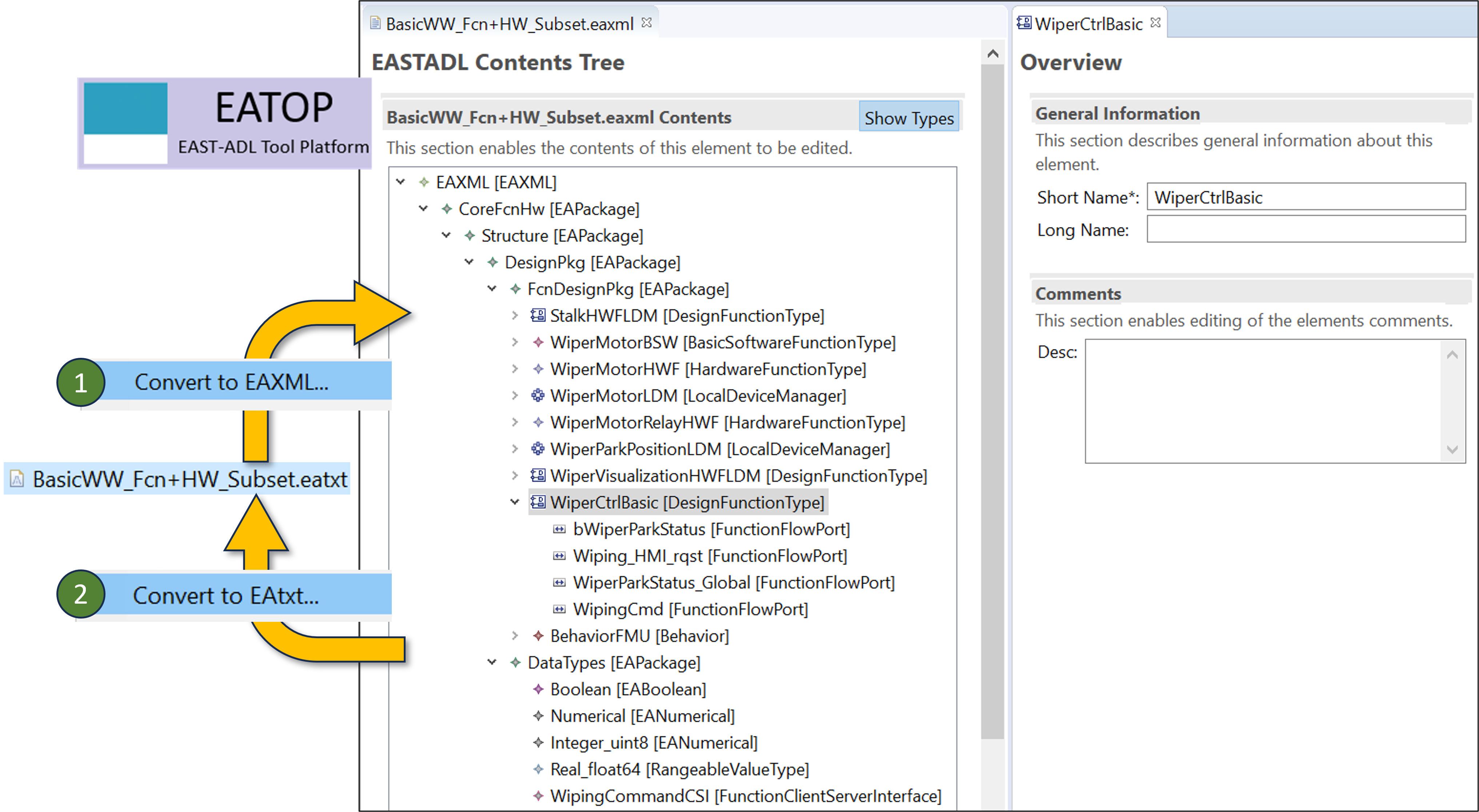}
    \caption{Serialization to EAXML}
    \label{fig:Serialization_example}
\end{figure}

In blended modeling with textual and other notations, serialization is a significant issue:
models with different notations adhere to different concrete syntaxes, e.g., the textual instances of EATXT adhere to the concrete syntax described by the EATXT grammar while the EATOP models adhere to a tree-based concrete syntax, however they adhere to the same abstract syntax.
Seamless blended modeling requires that models stored in one representation can be serialized in the other representation in a fully automated way.


\Cref{fig:Serialization_example} illustrates the workflow for a roundtrip between EATXT and EATOP.
As part of the EATXT editor, we provide actions to convert between the corresponding persistence formats.
That is, on any .eatxt file, the user can execute the action \textsf{Convert to EAXML...} (cf. annotation \textsf{1} in \Cref{fig:Serialization_example}).
This action serializes the EATXT model contents to an EAXML file, which can be opened in EATOP (cf. right-hand side of \Cref{fig:Serialization_example}).
Likewise, the user can execute the action \textsf{Convert to EAtxt...} (cf. annotation \textsf{2} in \Cref{fig:Serialization_example}) to transform the EATOP model contents to EATXT.
We currently support this roundtrip functionality for EAST-ADL models in version 2.1.12.

\section{Solution}
\label{sec:impl}
We developed the EATXT editor in a standard way of working with Xtext: we first generated the grammar from EAST-ADL's metamodel, and then generated the Xtext artifacts from the grammar to form the EATXT editor.
In addition to the standard workflow, we did the following additional work:
\begin{itemize}
    \item We adapted the grammar to make it more complete, user-friendly, and easy to use.
    \item We extended the Xtext artifacts to generate a more powerful editor with the features described in Section~\ref{sec:example}.
    \item We adopted an incremental development approach. We first developed a preliminary version of EATXT with a simplified EAST-ADL metamodel. After we completed the above two developments on the preliminary version, we transplanted the development results to the full version of EAST-ADL.
\end{itemize}



\subsection{Architecture}

\begin{figure}[tb]
	\centering
		\includegraphics[width=\textwidth]{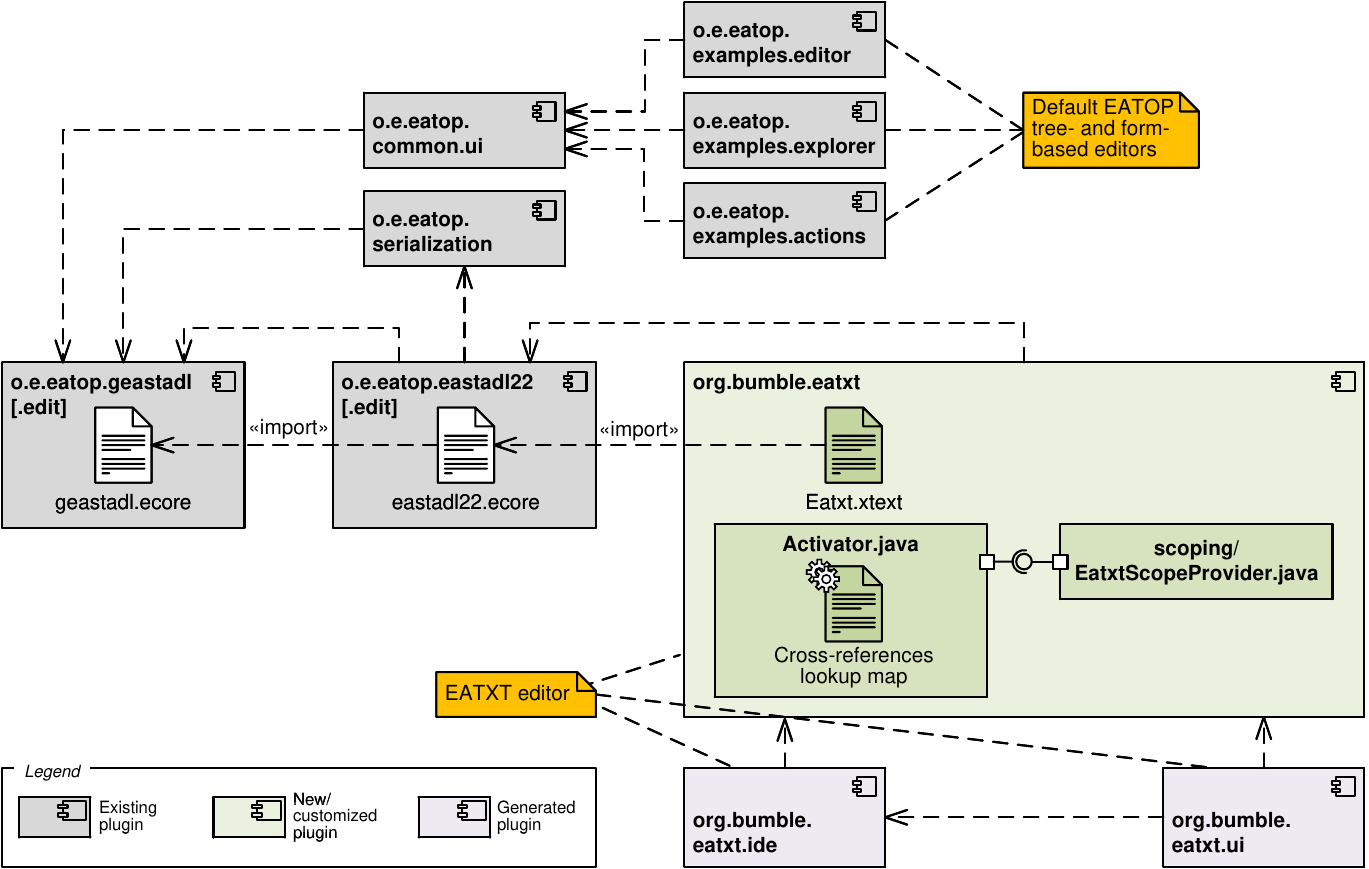}
	\caption{Architecture for the textual editor for EAST-ADL models and its relationships to EATOP.}
	\label{fig:architecture}
\end{figure}

\Cref{fig:architecture} depicts the architecture of EATXT from the perspective of modeling features, which includes the relationship between it and EATOP~\cite{Bumble2023Arch}. The EATXT editor contains different plug-ins, i.e., \textsf{org.bumble.eatxt}, \textsf{org.bumble.eatxt.simplified.ide}, and \textsf{org.bumble.eatxt.ui}, which are all generated by Xtext. In \Cref{fig:architecture}, the grammar \textsf{Eatxt.xtext} generated from the EAST-ADL metamodel (and also designed by us) as part of the plug-in \textsf{org.bumble.eatxt} imports EAST-ADL language concepts from \textsf{eastadl22.ecore} and associates them with the textual concrete syntax. The EATOP tree- and form-based editors are represented by the plugins \textsf{o.e.eatop.examples.[editor/explorer/actions]}. The customized (de-)serialization from/into EAXML is represented by the plugin \textsf{o.e.eatop.serialization}. The plugins \textsf{o.e.eatop.geastadl} and \textsf{o.e.eatop.eastadl22} provide the actual metamodels \textsf{geastadl.ecore} and \textsf{eastadl22.ecore}, respectively.

\subsection{Grammar Adaptation}
\label{grammardesign}
Following the relevant Xtext workflow \cite{Xtext}, we first generated a grammar from the available metamodel of EAST-ADL, and then adapted the generated grammar according to the specific requirements of our industrial partner regarding usability and grammar style. 
Our way of working with the industrial partner was iterative: in iterations, we discussed specific examples for chosen parts of the grammar, starting with a draft prepared by us that was refined during the discussion. Based on this, we adapted the grammar.
The adaptation included the following  main aspects: 
1) implementing missing primitive datatypes, e.g., datatype \texttt{Identifier}, 2) removing unnecessary elements, e.g., redundant curly braces, and 3) allowing empty elements.

\paragraph{Implement the incomplete primitive types.} In the metamodel of EAST-ADL, the instance type names of the primitive types including \texttt{Boolean}, \texttt{String}, \texttt{Integer}, etc. are specified as the primitive types of java, e.g., \texttt{Boolean} is specified as \texttt{java.lang.Boolean}. In the generated grammar, the definitions of these primitive types are missing, and Xtext places reminders in the form of ``implement this rule and an appropriate IValueConverter'' as code comments. Therefore, we implemented the definitions of the grammar rules for these primitive types. The most complex one is \texttt{Numerical}, which has different patterns such as binary, octal, decimal, etc. We implement these patterns separately and combined them with the ``or'' internal relationship as the definition of \texttt{Numerical}.



\paragraph{Remove unnecessary elements.} This presents an inherent issue when developing DSLs based on metamodels. That is, when generating grammar from metamodels using Xtext, Xtext often generates a lot of keywords, curly braces, commas, and other elements. However, a portion of these elements is unnecessary, leading users to input a lot of superfluous text and resulting in a program with a deep nesting structure, making the language less user-friendly and challenging to use. Listing~\ref{grammar_rule_eapackage} shows an example of a grammar rule from a generated grammar of EAST-ADL. When modeling a sub-package according to this grammar rule, the modeler needs to type the keyword ``subPackage'' and then curly brace, and then the keyword ``EAPackage'' and then curly braces which is unnecessary and cause a deep hierarchy. We employed the approach from~\cite{zhang2023creating} to remove unnecessary elements, with specific textual modifications on the grammar performed by the tool GrammarTransformer as described in~\cite{zhang2024supporting}. We retained the outermost curly braces (referred to as \emph{container braces}) for each grammar rule to accurately establish the hierarchical structure when serializing EATOP models to EATXT models. 

\begin{lstlisting}[caption={Excerpt of the generated grammar: the grammar rule \texttt{EAPackage}.}, label={grammar_rule_eapackage}]
EAPackage returns EAPackage:
    'EAPackage'
    '{'
        'shortName' shortName=Identifier
        ('category' category=Identifier)?
        ('uuid' uuid=String0)?
        ('name' name=String0)?
        ('ownedComment' '{' ownedComment+=Comment ( "," ownedComment+=Comment)* '}' )?
        ('subPackage' '{' subPackage+=EAPackage ( "," subPackage+=EAPackage)* '}' )?
        ('element' '{' element+=EAPackageableElement ( "," element+=EAPackageableElement)* '}' )?
    '}';
\end{lstlisting}

\paragraph{Allow empty elements.} In the generated grammar of EAST-ADL, the attributes contained in some grammar rules are optional, e.g., the grammar rule \texttt{EAPackage} shown in Listing~\ref{grammar_rule_eapackage}. However, curly braces are mandatory. Listing~\ref{example_of_mandatory_braces} shows a snippet in the domain program, which is an instance of the grammar rule \texttt{EAPackage}, and the identifier of this instance is \texttt{DesignPkg}. The default grammar forces the programmer to type curly braces even if she does not intend to provide instances for the attributes of grammar rule \texttt{EAPackage}. The curly braces are redundant in this case. Our solution is to add a set of symbols (i.e., \texttt{()?}) outside the container curly braces to make them optional, including the attributes enclosed by the container curly braces. Finally, in the same case, the user only needs to type \texttt{EAPackage DesignPkg} without the need to type braces.

\begin{lstlisting}[caption={Excerpt of domain program: an example of redundant and mandatory braces.}, label={example_of_mandatory_braces}]
EAPackage DesignPkg
{

}
\end{lstlisting}

\subsection{Content-assist}
\label{editorartifact}
Xtext provides a default content-assist feature, but it has the following three problems. First, it offers inaccurate proposals; second, its proposals do not display the full path name of the element; and third, the performance is slow. The \textsf{EatxtScopeProvider} can solve the second problem mentioned above. In addition, after obtaining the scope collection containing the fully qualified names of all elements in instances, we filter and calculate proposals for the target location through type judgment. These proposals are the text that can be entered at the location, thereby solving the first problem. 



The EAST-ADL metamodel is quite complex and encompasses 291 metaclasses (cf.~\Cref{sec:background_EASTADL}). 
Consequently, we discovered that the content-assist suffers from performance issues, because Xtext in its default implementation completely traverses the large and complex EAST-ADL metamodel on every keystroke triggering the content-assist. 
This is the cause of the third problem. Thus, we implemented a caching feature that computes and stores all cross-references in an in-memory lookup map on the EATXT editor startup \cite{xtextautomation2023}.
This lookup map strongly improves the content-assist performance.

In \Cref{fig:architecture}, the lookup map is visualized as \textsf{Cross-references lookup map}. The \textsf{lookup map} is populated in the \textsf{Activator} when the plugin starts. Whenever a scope is required, the \textsf{EatxtScopeProvider} just retrieves the pre-filled map from the \textsf{Activator}.


\subsection{Serialization}

To implement serialization and de-serialization between the EATXT and EAXML formats, we used the serialization features provided by EMF, Xtext, and particularly EATOP. Specifically, when serializing an EATXT model in EAXML format, we first obtain the EATXT resource and convert it into an in-memory model. Then we configure \texttt{XMLResource} to ensure that the model can be serialized into XML format. Then we save the model as an EAXML model. 
Similarly, we implemented the features of de-serialization from EAXML models to EATXT models in the opposite way. However, three customizations were necessary and implemented by us, and the first two are in the grammar. First, we retained the outermost curly braces of all grammar rules so that Xtext's automatic formatting could put different elements on different lines, thus achieving the basic format of the textual model. Second, the type \texttt{UUID} in EAST-ADL cannot be resolved, so we designed a terminal UUID (see the supplementary material). Third, some attributes in the EAXML model may have empty values. An error will be reported if an attribute with an empty value is de-serialized to EATXT. Therefore, we skipped the empty attributes when writing the EATXT model file, i.e., all attributes shown in EATXT contain values.
Beyond these aspects, we simply trigger the EATOP serialization encapsulated in the plug-in \textsf{o.e.eatop.serialization}, which is part of the architecture in \Cref{fig:architecture}.

\section{Previous use of EATXT in research}
\label{sec:previous_use}
EAXT has been used as a case language in several previous research efforts.

In~\cite{zhang2024supporting}, we present the technique GrammarTransformer for supporting the automated transformation of grammars in a metamodel-driven language development scenario, to reduce manual effort for updating grammars after changes to the corresponding metamodel.
EATXT was used as one of the case languages to evaluate GrammarTransformer. 
Specifically, we considered the EATXT counterparts to both a simplified and the full version of EAST-ADL  with 91 and 291 metamodel classes, respectively.
We first focused on the simplified version. Using  GrammarTransformer, we derived a grammar from the simplified EAST-ADL metamodel containing 91 grammar rules. To adapt the generated grammar to our industry partner's requirements,  more than 500 lines of text in 70 grammar rules had to be adapted, if done manually. However, we used GrammarTransformer to complete this modification with a configuration containing only 22 adaptation rules.
After that, we considered the full EAST-ADL metamodel, for which the generated grammar contained 291 grammar rules. To adapt the grammar to the expert-created grammar, we needed to modify 233 of the grammar rules, which involved modifying more than 2,000 lines of text. We completed this adaptation by modifying only ten adaptation rules in the rule configuration that we created for the simplified version. 
These results show that GrammarTransformer can effectively support grammar adaptation during metamodel-based DSL evolution.

Manually configuring the adaptation rules of GrammarTransformer is sometimes cumbersome, so in~\cite{zhang2023automated}, we developed a tool called ConfigGenerator to automatically generate the configuration of adaptation rules. The specific rationale is that there are differences between the generated grammar and the expert-created grammar, e.g., the same grammar rule differs by one keyword in the two grammars. ConfigGenerator can analyze and collect the differences between the two grammars, and find specific adaptation rules from the adaptation rule list of GrammarTransformer to adapt each difference. EATXT is one of the case languages in this study. ConfigGenerator successfully compared the grammar generated from the EAST-ADL metamodel and the expert-created grammar of EATXT, collected the difference between them, and extracted a configuration containing multiple adaptation rules based on the comparison. Finally, this configuration successfully drove GrammarTransformer to adapt the generated grammar to the expert-created grammar of EATXT. The experimental results showed that ConfigGenerator could reduce the effort of using GrammarTransformer in a way that extracts adaptation rule configurations.

When generating textual editors for large and highly structured metamodels, it is possible to extend Xtext’s generator capabilities and the default implementations it provides. These extensions provide additional features such as formatters and more precise scoping for cross-references. However, for large metamodels in particular, the realization of such extensions typically is a time-consuming, awkward, and repetitive task. For some of these tasks, in~\cite{xtextautomation2023}, we motivated, presented, and discussed automatic solutions that exploit the underlying metamodel structure. Furthermore, we showed how the metamodel structure can be used in text editor development in the context of EATXT.

\section{Conclusion and Future Work}
\label{sec:conclusion}
In this paper, we introduce the concrete textual syntax and editor EATXT for EAST-ADL textual modeling. EATXT serves as an exemplar of an industry-level open-source textual syntax and editor that can be used to support the development of language engineering approaches in the context of blended modeling.
We briefly outline our motivation for developing it, its main features, and implementation aspects. We also provide an illustrative example. As mentioned in the introduction, developing EATXT is an incremental process. 

As for future work, we plan to make EATXT support different versions of the EAST-ADL schema at the same time. In addition, we also plan to integrate EATXT into EATOP and make it a part of EATOP software.

\bibliographystyle{abbrv}

\bibliography{main}

\begin{thebibliography}{10}

\bibitem{Xtext}
L.~Bettini.
\newblock {\em Implementing domain-specific languages with Xtext and Xtend}.
\newblock Packt Publishing Ltd, 2016.

\bibitem{ciccozzi2019blended}
F.~Ciccozzi, M.~Tichy, H.~Vangheluwe, and D.~Weyns.
\newblock Blended modelling-what, why and how.
\newblock In {\em 2019 ACM/IEEE 22nd International Conference on Model Driven Engineering Languages and Systems Companion (MODELS-C)}, pages 425--430. IEEE, 2019.

\bibitem{cuenot2007managing}
P.~Cuenot, D.~Chen, S.~Gerard, H.~Lonn, M.-O. Reiser, D.~Servat, C.-J. Sjostedt, R.~T. Kolagari, M.~Torngren, and M.~Weber.
\newblock Managing complexity of automotive electronics using the east-adl.
\newblock In {\em 12th ieee international conference on engineering complex computer systems (iceccs 2007)}, pages 353--358. IEEE, 2007.

\bibitem{damasceno2021quality}
C.~D.~N. Damasceno and D.~Str{\"u}ber.
\newblock Quality guidelines for research artifacts in model-driven engineering.
\newblock In {\em 2021 ACM/IEEE 24th International Conference on Model Driven Engineering Languages and Systems (MODELS)}, pages 285--296. IEEE, 2021.

\bibitem{david2023blended}
I.~David, M.~Latifaj, J.~Pietron, W.~Zhang, F.~Ciccozzi, I.~Malavolta, A.~Raschke, J.-P. Stegh{\"o}fer, and R.~Hebig.
\newblock Blended modeling in commercial and open-source model-driven software engineering tools: A systematic study.
\newblock {\em Software and Systems Modeling}, 22(1):415--447, 2023.

\bibitem{eadl}
{EAST-ADL Association}.
\newblock {EAST-ADL}, 2024.
\newblock \url{https://www.east-adl.info/}, last accessed June 2024.

\bibitem{eatop}
{Eclipse Foundation}.
\newblock {Eclipse EATOP}, 2024.
\newblock \url{https://projects.eclipse.org/projects/modeling.eatop}, last accessed June 2024.

\bibitem{mwe}
{Eclipse Foundation}.
\newblock {Eclipse Modeling Workflow Engine}, 2024.
\newblock \url{https://projects.eclipse.org/projects/modeling.emf.mwe}, last accessed June 2024.

\bibitem{behrens2008xtext}
{Eclipse Foundation}.
\newblock Xtext user guide, 2024.
\newblock \url{https://eclipse.dev/Xtext/documentation/}, last accessed June 2024.

\bibitem{erdweg2015evaluating}
S.~Erdweg et~al.
\newblock {Evaluating and comparing language workbenches: Existing results and benchmarks for the future}.
\newblock {\em Comput. Lang. Syst. Struct.}, 44:24--47, 2015.

\bibitem{xtextautomation2023}
J.~Holtmann., J.~Steghöfer., and W.~Zhang.
\newblock Exploiting meta-model structures in the generation of xtext editors.
\newblock In {\em Proceedings of the 11th International Conference on Model-Based Software and Systems Engineering - MODELSWARD,}, pages 218--225. SciTePress, 2023.

\bibitem{Bumble2023Arch}
{ITEA 4}.
\newblock {BUMBLE Deliverable D3.1\,---\,Architecture Description for BUMBLE Eclipse Platforms} (version 3), 2023.
\newblock \url{https://itea4.org/project/workpackage/document/download/8690/BUMBLE_D3.1v3_Final.pdf}, last accessed June 2024.

\bibitem{yuan2014safety}
W.~Yuan.
\newblock Safety analysis and model transforming in east-adl.
\newblock 2014.

\bibitem{zhang2023creating}
W.~Zhang, R.~Hebig, J.-P. Stegh{\"o}fer, and J.~Holtmann.
\newblock Creating python-style domain specific languages: A semi-automated approach and intermediate results.
\newblock In {\em MODELSWARD}, pages 210--217, 2023.

\bibitem{zhang2023automated}
W.~Zhang, R.~Hebig, D.~Str{\"u}ber, and J.-P. Stegh{\"o}fer.
\newblock Automated extraction of grammar optimization rule configurations for metamodel-grammar co-evolution.
\newblock In {\em Proceedings of the 16th ACM SIGPLAN International Conference on Software Language Engineering}, pages 84--96, 2023.

\bibitem{eatxtarchive}
W.~Zhang, J.~Holtmann, and D.~Str{\"u}ber.
\newblock Dataset for {'EATXT: A textual concrete syntax for EAST-ADL'}, 2024.
\newblock \url{https://osf.io/gkdqw/?view_only=0764537e09584301a189c94210813b8b}, last accessed June 2024.

\bibitem{zhang2024supporting}
W.~Zhang, J.~Holtmann, D.~Str{\"u}ber, R.~Hebig, and J.-P. Stegh{\"o}fer.
\newblock Supporting meta-model-based language evolution and rapid prototyping with automated grammar transformation.
\newblock {\em Journal of Systems and Software}, page 112069, 2024.

\end{thebibliography}

\end{document}